\documentclass[numbib]{comnet}
\usepackage[utf8]{inputenc}
\usepackage{graphicx}
\usepackage[T1]{fontenc}

\begin{document}

\title{Universal features of mountain ridge networks on Earth}

\shorttitle{Universal features...}
\shortauthorlist{R. Rak {\it et al.}}

\author{%
{\sc Rafał Rak}\\[2pt]
Faculty of Mathematics and Natural Sciences, University of Rzeszów, ul. Pigonia 1, 35-310 Rzeszów\\
{rafalrak@ur.edu.pl}\\
Complex Systems Theory Department, Institute of Nuclear Physics, Polish Academy of Sciences, ul. Radzikowskiego 152, 31-342 Kraków, Poland\\[6pt]
{\sc Jarosław Kwapień}$^{*}$\\[2pt]
Complex Systems Theory Department, Institute of Nuclear Physics, Polish Academy of Sciences, ul. Radzikowskiego 152, 31-342 Kraków, Poland\\
{Corresponding author: jaroslaw.kwapien@ifj.edu.pl}\\[6pt]
{\sc Paweł Oświ\c ecimka}\\[2pt]
Complex Systems Theory Department, Institute of Nuclear Physics, Polish Academy of Sciences, ul. Radzikowskiego 152, 31-342 Kraków, Poland\\
{pawel.oswiecimka@ifj.edu.pl}\\[6pt]
{\sc Paweł Zi\c eba}\\[2pt]
Energy Business Intelligence Systems, Piłsudskiego 32, 35-001 Rzeszów, Poland\\
{pawel.zieba@interia.pl}\\[6pt]
{\sc and}\\[6pt]
{\sc Stanisław Drożdż}\\[2pt]
Complex Systems Theory Department, Institute of Nuclear Physics, Polish Academy of Sciences, ul. Radzikowskiego 152, 31-342 Kraków, Poland\\
Faculty of Physics, Mathematics and Computer Science, Cracow University of Technology, ul. Podchorążych 1, 30-084 Kraków, Poland\\
{stanislaw.drozdz@ifj.edu.pl}}

\maketitle

\begin{abstract}
{Compared to the heavily studied surface drainage systems, the mountain ridge systems have been a subject of less attention even on the empirical level, despite the fact that their structure is richer. To reduce this deficiency, we analyze different mountain ranges by means of a network approach and grasp some essential features of the ridge branching structure. We also employ a fractal analysis as it is especially suitable for describing properties of rough objects and surfaces. As our approach differs from typical analyses that are carried out in geophysics, we believe that it can initialize a research direction that will allow to shed more light on the processes that are responsible for landscape formation and will contribute to the network theory by indicating a need for the construction of new models of the network growth as no existing model can properly describe the ridge formation. We also believe that certain features of our study can offer help in the cartographic generalization.

Specifically, we study structure of the ridge networks based on the empirical elevation data collected by Shuttle Radar Topography Mission. We consider mountain ranges from different geological periods and geographical locations. For each mountain range, we construct a simple ``topographic'' network representation (i.e., the ridge junctions are nodes) and a ``ridge'' representation (i.e., the ridges are nodes and the junctions are edges) and calculate the parameters characterizing their topology. We observe that the topographic networks inherit the fractal structure of the mountain ranges but do not show any other complex features. In contrast, the ridge networks, while lacking the proper fractality, reveal the power-law degree distributions with the exponent $1.6 \le \beta \le 1.7$. By taking into account the fact that the analyzed mountains differ in many properties, these values seem to be universal for the earthly mountainous terrain.}
{Complex networks, Mountain ridges, Scale-free networks, Fractal networks, Universality.}
\end{abstract}

\section{Introduction}

Mountains are among the most common geomorphological structures~\cite{stahr2015}. Sometimes they assume a simple form of isolated peaks like the active stratovolcanos or the tepuis, but typically they form whole ranges with a rich internal structure of ridges that form chains of peaks, bifurcate and send side ridges, and disappear in valleys or lowlands. Their complexity stems from the fact that they are subject to the concurrent yet interwoven processes of upthrust, folding, rock-mass displacement along faults, weathering, fluvial and glacial erosion and transport, and the downhill mass movements due to landslides and diffusion, in which an originally flatter but elevated terrain is gradually carved first into shallow rills and then shaped to a combination of more prominent ridges, gullies, and valleys~\cite{koyama1997,babault2012,hovius1998,korup2007,%
pelletier2003,mcguire2013,howard1994,montgomery2002,molnar2007}. Since the valleys are associated with the surface drainage systems, except for the rather seldom occurring endorheic areas, an overwhelming majority of the mountain ranges form the hierarchical, tree-like ridge systems complementary to the river networks~\cite{werner1988}.

A ridge-valley system of a typical mountain range irrespective of its geographical location assumes a dendritic, parallel or trellis-like structure with a clear hierarchy and self-similarity. It sometimes shows also a considerable amount of regularity that is expressed, e.g., in evenly-spaced parallel ridges/valleys on different hierarchy levels, which amplifies the overall impression of self-similarity~\cite{perron2009}. These features put the ridge and drainage systems among the well-recognizable examples of natural fractals. The structural self-similarity has extensively been studied in the drainage systems, for which there exist several well-established power-law relations that functionally link such quantities as stream length and stream basin length, stream length and basin area, and so on~\cite{tarboton1988,dodds1999,labarbera1989}. Probability distributions of some of those quantities, like basin area and stream length also reveal power-law tails. If one reduces mountain ridges to ridge lines, it occurs indeed that their structure mirror to some extent the drainage networks on the same territory (see~\cite{werner1988} for rigorous theoretic arguments), so one may expect intuitively that the analoguous power-law relations are valid for the ridges.

There is a number of ways to map ridge and valley systems into graph representations. The main related requirement is that such networks have to preserve specific features of the systems' topology, like, e.g., ridge bifurcations, valley junctions, peaks, cols, or pits, depending on a studied problem. Among the most extensively studied examples of such representations one can mention the Reeb graphs~\cite{reeb1946}, the contour trees~\cite{morse1968}, the Pfaltz graphs~\cite{pfaltz1976}, and the surface networks~\cite{mark1977}. While each of these representations uses its own characteristic mapping, none of them considers the ridge systems in the simplest manner as the connected ridge lines without any internal structure consisting of the distinguished critical points: peaks and passes. These points can be rather irrelevant to an analysis of the ridge-line systems. Moreover, some of the above representations are defined in such a way that they interweave the ridges with the valleys (or the river channels) and the resulting graphs do not distinguish among those objects. Thus, from a viewpoint of the sole ridge-line analysis, by constructing the Reeb graphs or the surface networks, one faces a substantial information overload and superfluous contamination. Even if that could be filtered out, some method-specific problems can remain (like, e.g., dealing with the monkey saddles).

Therefore, if one is interested solely in the ridge-line system topology, none of the above representations seems to be the most natural and effective. Instead, we choose an approach that uses two following representations. We call the first one a ``topographic'' representation. Here, the ridges are represented by edges and the ridge bifurcations are represented by nodes. Both the binary and the weighted networks can be constructed in this manner with the weights defined by heights (either maximal, average, or relative) of the corresponding ridge segments. The most characteristic property of such networks is their direct visual correspondence to the topography of the areas studied (expressed by, e.g., their ridge maps), especially if the edge lengths are proportional to the respective segment lengths. The other network representation (called a ``ridge'' representation) is relatively more abstract from the topographic point of view: each node is associated with a ridge and edges connect this node with the nodes representing the parent ridge and the side-ridges. In this case constructing the weighted network is also conceivable with the weights reflecting the bifurcation point heights.  Obviously, the networks in both representations form trees if the corresponding mountain range does not contain any endorheic territory. The dendritic structure of the resulting networks is a substantial advantage of our approach over, for instance, the cycle-abundant surface networks, since some basic topological parameters (like the average shortest path length) are much easier to be calculated in this case.

We shall address two questions: (1) what are the topological characteristics of the networks in both representations? and (2) are these characteristics range-dependent or are there any universal topological features of the networks that can reflect some universality of the mountainous terrain structure? The most interesting problem related is whether its self-similarity leaves any marks on the network topology.

As besides self-similarity mountains exemplify also another essential terrain feature - roughness, it is natural to describe and quantify their structure by employing the fractal analysis. We follow this path predominantly on a level of the network stucture, which distinguishes our work among the fractal landscape analyses documented in literature so far.

\section{Data processing and network construction}

\subsection{Ridge-axis detection}

Our analysis is based on the empirical terrain elevation data obtained by the 2000 Shuttle Radar Topography Mission (SRTM)~\cite{srtm}, which for many selected mountainous areas is available from a dedicated internet site~\cite{viewfinder}. The resolution of the data used is 3'' in both the meridional and the zonal direction, which is equivalent to $\approx 90$ m along a meridian and $\approx 90 {\rm cos}\phi$ m along a latitude circle $\phi$. This resolution seems to be sufficient for creating the ridge maps since it is better than typical separation of the neighbouring ridges. The data is provided as a set of rectangle arrays of size 1201$\times$1201 points (1$\times$1 arc degree). Each data point on a grid represents elevation of the corresponding terrain point. The number of such arrays to be processed depends on a mountain group and varies from 11 to 306 (Tab.~\ref{tab1}). The ridges are approximated by their axes being the series of points where the terrain slopes on both opposite sides. 

There are many available methods of extracting the exact run of the ridge axes (direct or indirect via extracting the drainage basin borders)~\cite{collins1975,peucker1975,ocallaghan1984,%
jenson1988,tarboton1997,wood1998,schneider2005,freitas2016}. They differ in precision and computer resource requirements, but here we simply need a method that produces few errors in the ridge-line detection and requires moderate computer resources. In order to achieve this, we use a modified version of the Profile Recognition and Polygon Breaking Algorithm (PPA)~\cite{chang1998,chang2007,zhou2007}. This choice is based on our previous experience indicating the method's satisfactory output precision and the already-developed software~\cite{glowacki2016}, while it might not necessarily be the optimal method in terms of the combined precision and computing resources used. However, as the ridge-line extraction is only a data-preparation step for the main part of our study, we decided to apply the PPA method, provided the obtained ridge-line courses agree with the terrain topography (which is the case here, indeed).

\begin{center}
\begin{table}[h!]
\begin{footnotesize}
\begin{tabular}
{|l|c|c|l|c|}
\hline
\multicolumn{1}{|c|}{Range} & Area (km$^2$) & Height (m) & \multicolumn{1}{|c|}{Origin} & No. data pts. \\
\hline\hline
Alps & 207,000 & 4,807 & Orogenic & 36,895,401 \\
\hline
Baetic Mountains & 100,000 & 3,478 & Orogenic & 11,255,768 \\
\hline
Pyrenees & 19,000 & 3,404 & Orogenic & 10,332,333 \\
\hline
Scandinavian Mountains & 243,000 & 2,469 & Non-orogenic & 138,497,654 \\
\hline
Himalayas & 594,400 & 8,848 & Orogenic & 40,295,462$^{*}$ \\
\hline
Southern Alps & 36,700 & 3,724 & Orogenic & 23,619,193 \\
\hline
Appalachian Mountains & 531,000 & 2,037 & Orogenic & 208,854,395 \\
\hline
Atlas Mountains & 775,340 & 4,167 & Orogenic & 240,495,193 \\
\hline
Andes & 3,371,000 & 6,961 & Orogenic & 320,431,192$^{*}$ \\
\hline
\end{tabular}
\caption{Characteristics of the mountain ranges studied in the present work. (*) Only parts of the Himalayas and the Andes were studied.}
\end{footnotesize}
\label{tab1}
\end{table}
\end{center}

\begin{figure}[h]
\includegraphics[width=0.9\textwidth]{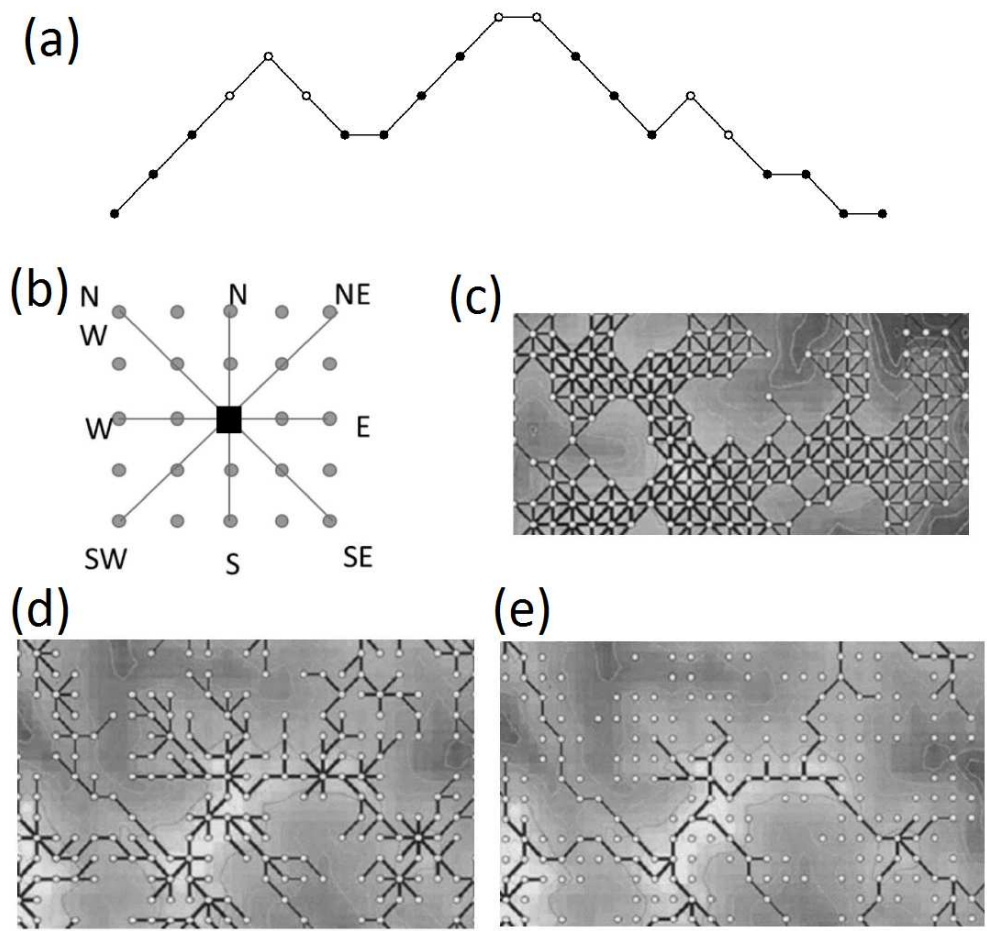}
\caption{(a) A sample line profile of the SRTM data illustrating how the candidate points for a ridge axis (open circles) are identified. (b) All possible 5-point profiles containing the same central SRTM point. (c) Candidate points connected by line segments on a piece of the SRTM grid after inspecting all possible cross-sections. (d) The effect of applying the MST filter. (e) The ridge axes after the final step of the branch reduction.}
\label{fig1}
\end{figure}

PPA consists of several steps. First, by using the moving $p$-point line profiles, one obtains a set of straight-line cross-sections of the elevation grid (Fig.~\ref{fig1}(a)), in which each cross-section has one of 4 orientations: N-S, E-W, NE-SW, and NW-SE (Fig.~\ref{fig1}(b)). We used profiles of length $p=5$ following Ref.~\cite{chang2007}, which is optimal as it allows us both to preserve continuity of the ridge axes (by surpassing the noise effects) and to reduce the computation burden~\cite{glowacki2016}. In each position of the moving profile, its central point is tested for being a ridge-axis candidate. If on both sides of this point there is at least one point that has lower elevation than the central point's one, this central point is considered to be such a candidate (Fig.~\ref{fig1}(a)). The candidate points that are neighbours on a cross-section are then connected by a line segment (Fig.~\ref{fig1}(c)).

As the resulting structure is overpopulated by intersecting lines, the second step consists of reducing the number of the line segments by the following actions: (1) eliminating the diagonal segment with a lower average elevation if two such diagonal segments intersect in a grid square, and (2) eliminating those of the two or three neighbouring parallel segments that have lower average elevation and leaving only the one with the highest average elevation (if such parallel segments exist, the segment left is called a reliable segment). The average elevation of a segment is calculated based on elevations of its ends. However, after this step, the structure still has too many segments and cycles that do not reflect any real ridges.

Typically, the so-called polygon breaking algorithm is applied at this phase~\cite{chang1998}. It consists of sorting the segments in ascending order according to their average elevation and, by starting from the lowest one, successively removing those segments that are parts of closed polygons. As this step requires much computer resources, certain workarounds are used in order to speed up computation, like, e.g., the dead-end detection that prevents the algorithm from considering those segments that have already been identified as not containing cycles~\cite{chang2007}. However, even with those workarounds, the polygon-breaking algorithm significantly bounds the maximum possible region size to be studied and for such mountains as the Alps it demands to allocate too much amount of CPU time even on a supercomputer.

However, a much more efficient minimum spanning tree approach can be applied~\cite{bangay2010}. In this approach, the grid of elevation points connected by the line segments, being the output of the profile recognition algorithm, is considered a weighted network, in which the points are the nodes, the segments are the edges. For such a network spanned by $N$ nodes, the minimum spanning tree (MST) is a network subset that contains all the nodes, but only $N-1$ edges selected in such a way that the total sum of edge weights is minimum possible~\cite{barrow1985,mantegna1999}. As mountain ridges are convex structures, in order for them to form an MST, we have to transform elevation $h(x,y)$ into depth $d(x,y)=H-h(x,y)$ relative to some reference level $H$, selected in such a way that $d(x,y)>0$ for all the grid points with the coordinates $x,y$. This transformation preserves the metric properties of $h$ and $d$. Now we consider the segment average depths as the weights of the corresponding network edges. Out of several different MST-construction algorithms, we choose the Prim's algorithm~\cite{prim1957}. We sort the weights of a network in ascending order and connect the first pair of nodes by the edge with the smallest weight. Then, by considering edges with higher weights one by one, we connect only those node pairs, in which at most one node has already been connected. We proceed with this procedure till there is no unconnected node left.

\begin{figure}
\includegraphics[width=0.5\textwidth]{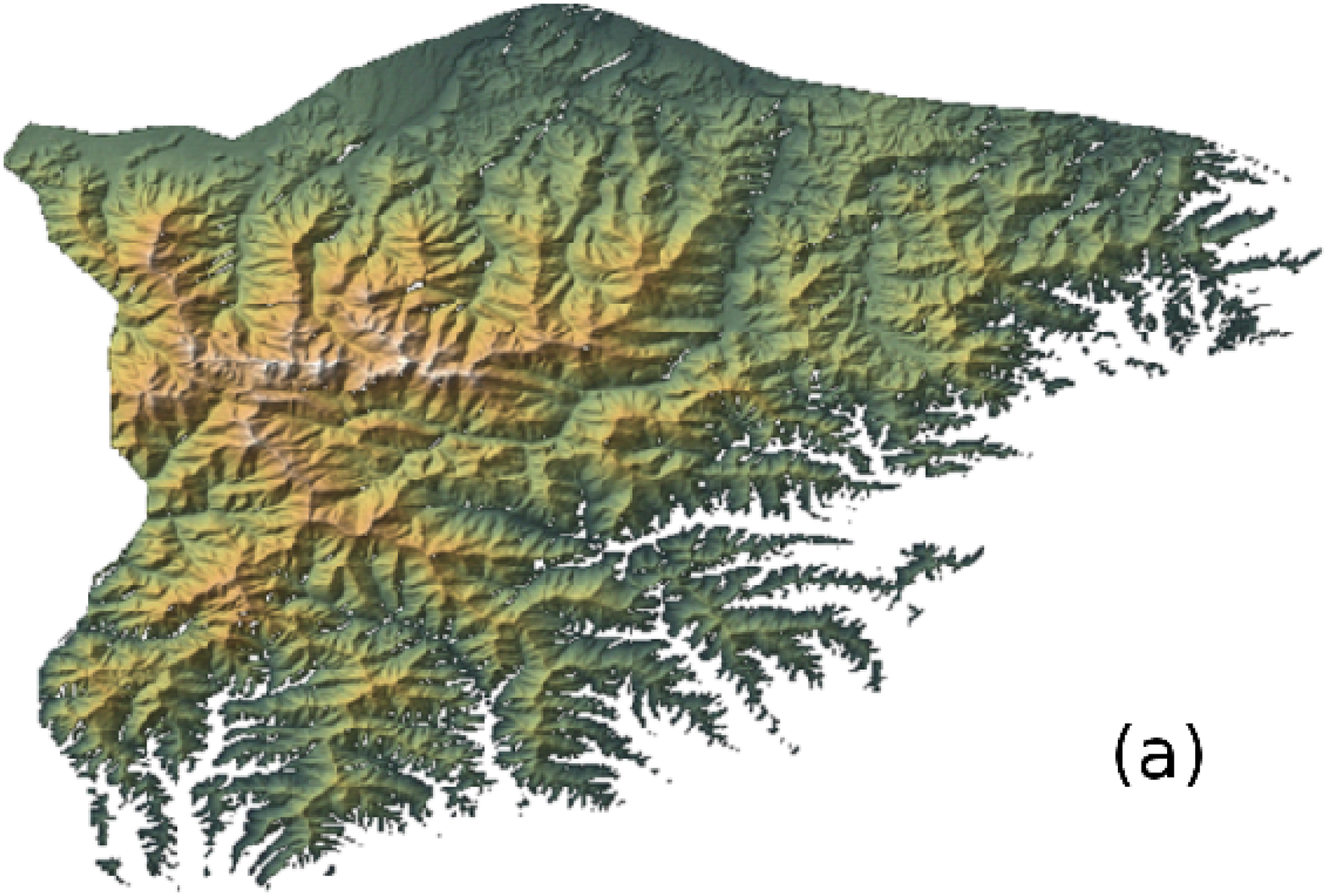}\includegraphics[width=0.5\textwidth]{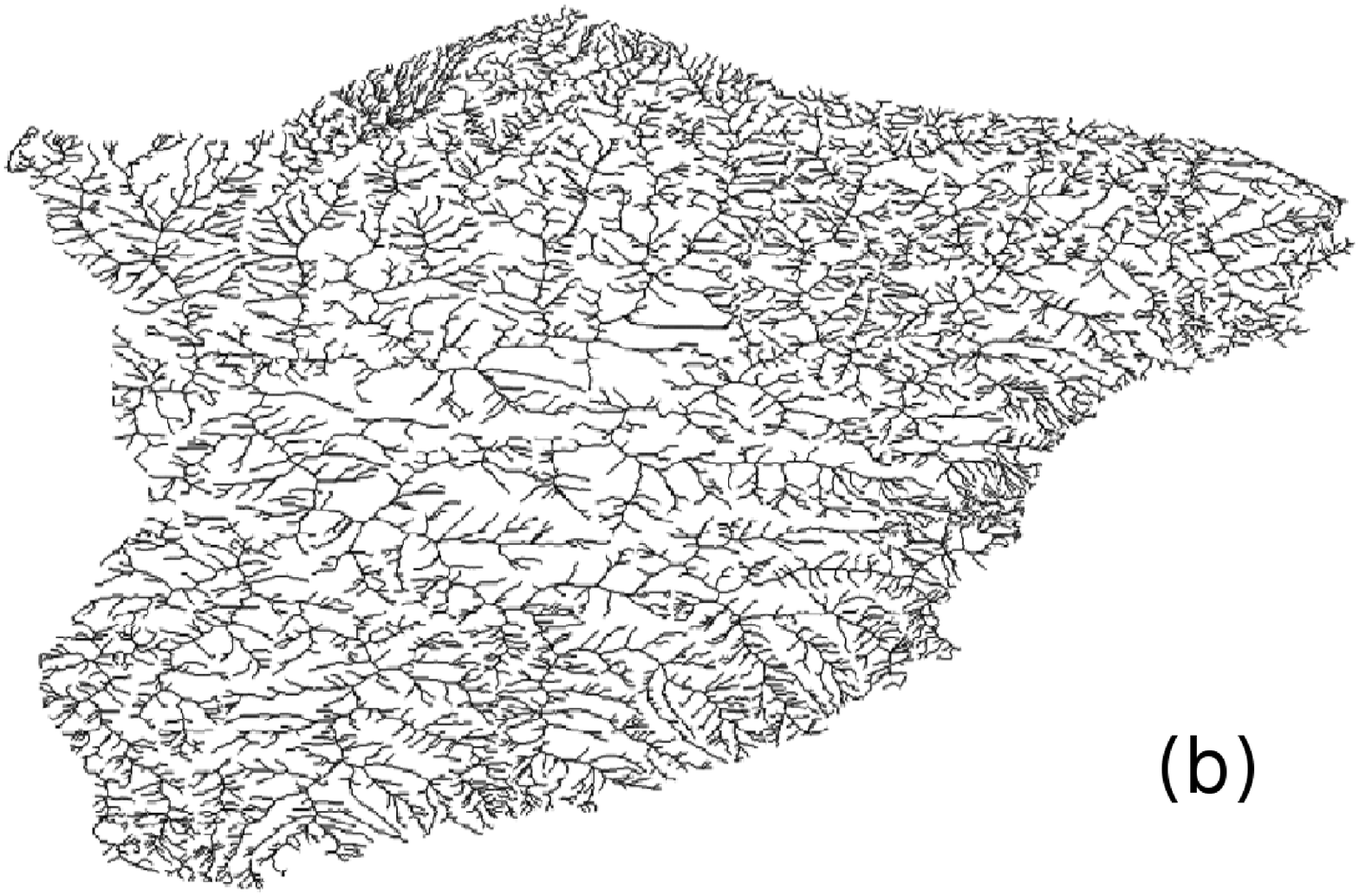}
\caption{(a) A shaded relief map of the Ligurian Alps created from the 3'' SRTM data. (b) The ridge map obtained by means of the ridge-axis detection procedure described in Section 2.}
\label{fig2}
\end{figure}

The output of the MST procedure is a tree consisting of branches representing the run of the ridge axes (Fig.~\ref{fig1}(d)). However, its structure may still contain spurious double or multiple parallel branches representing in fact the same ridge. It may also contain short branches that are associated with terrain roughness rather than with any real ridge. Therefore, a filtering method called Branch Reduction has to be applied at this step~\cite{chang2007}. All the branches are shortened in order for them to be terminated only with a reliable segment and the branches shorter than half a profile length ($<p/2$) are removed together with the segments that are not connected to any larger structure (Fig.~\ref{fig1}(e)). We verified carefully on sample maps covering pieces of a few mountain ranges that the ridge-axis trees, after the branch reduction step, coincide with the actual ridge structure of those mountains (as seen in topographic contour maps) with satisfactory precision (i.e., without any meaningful impact on the outcomes of our statistical analysis). In Fig.~\ref{fig2} we show the result of the above ridge-axis-detection procedure applied to the Ligurian Alps.

(Each piece of software is available upon request from one of the authors (RR).)

\subsection{Network construction}

In order to prepare data for our study, we selected several mountain ranges (the Alps, the Pyrenees, the Baetic Mountains, the Scandinavian Mountains, the Southern Alps, the Appalachian Mountains, the Atlas Mountains, and parts of the Himalayas and the Andes) and obtained the SRTM data for the corresponding geographical regions. The mountains of interest differ in maximum height and size of the occupied area as well as they come from different geological eras. All ranges except one are products of tectonic plate collisions (orogeny), while the Scandinavian Mountains, being a part of the continental passive margin, are believed to originate solely due to the uplift and erosion of the interface between the continental and oceanic litosphere~\cite{japsen2000,chalmers2010} (see Tab.~\ref{tab1}).

\begin{figure}
\begin{center}
\includegraphics[width=0.6\textwidth]{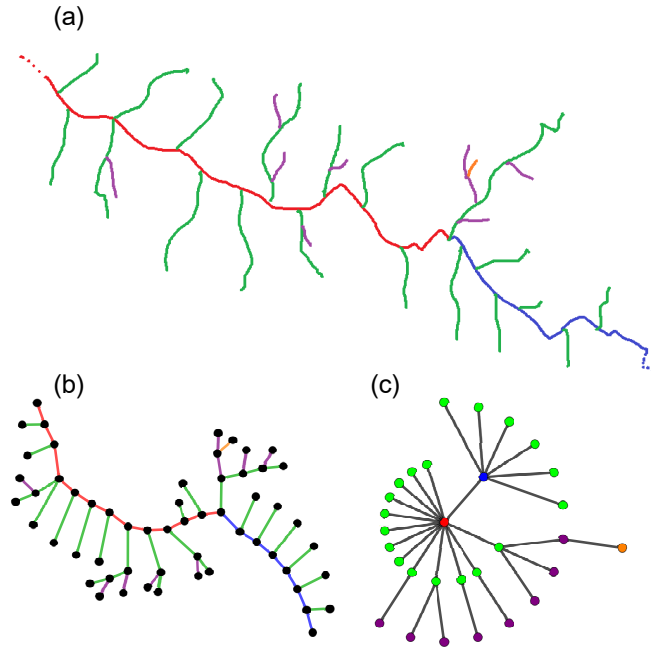}
\end{center}
\caption{(a) A ridge-line map of a hypothetical ridge system with colours denoting the ridge orders (with an exception for the blue and the red line both denoting parts of the main ridge) together with (b) a topographic network and (c) a ridge network based on the map (a).}
\label{fig3}
\end{figure}

\begin{figure}
\includegraphics[width=1.0\textwidth]{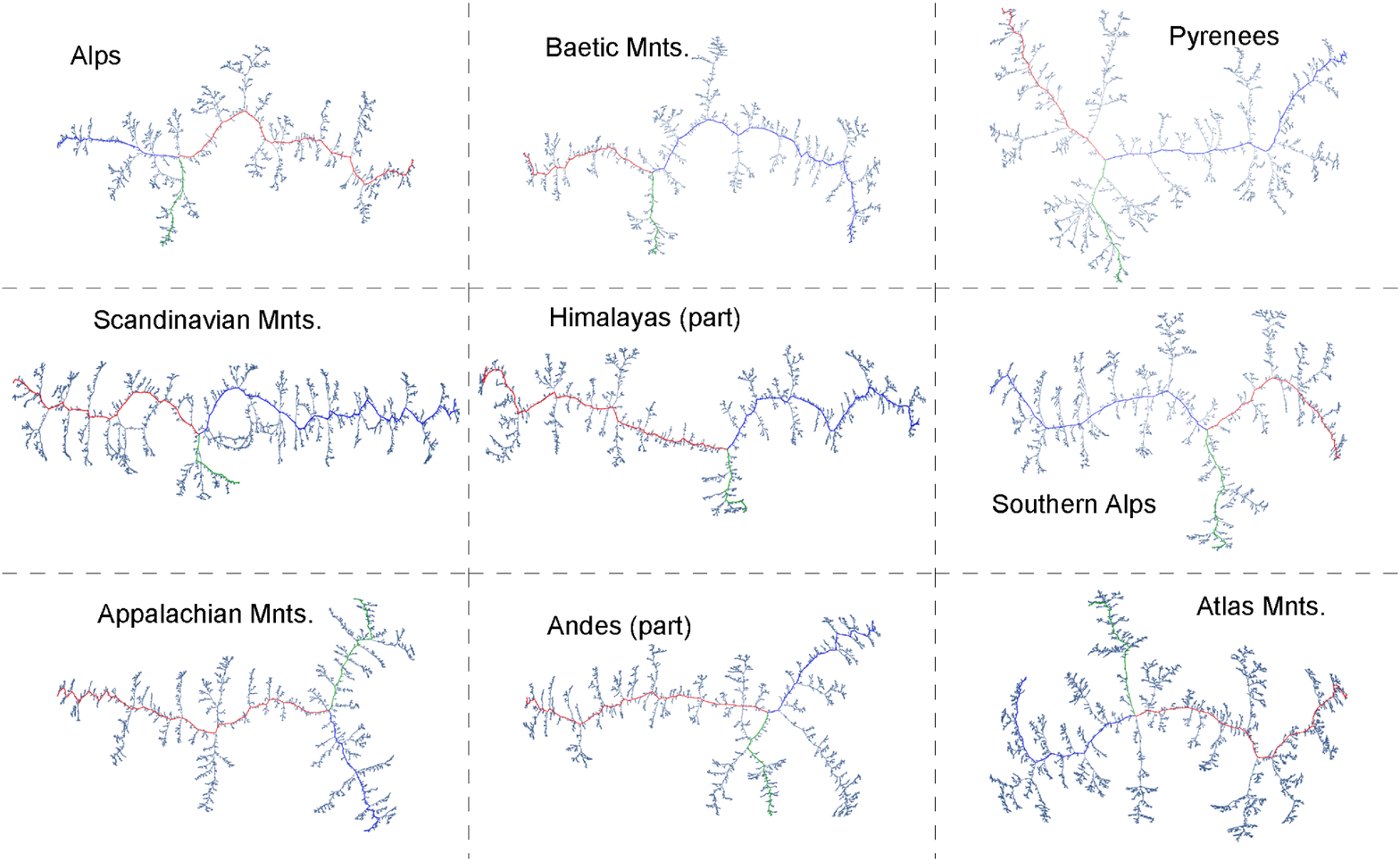}
\caption{Topographic networks representing the mountain ranges considered in this work. Each node corresponds to a ridge junction point or a ridge end point, while each edge connects a pair of nodes such that the respective terrain points are connected by a ridge axis without passing through any intermediate node. The edges are binary and their lengths are unrelated to the corresponding ridge-segment lengths. In each case the main ridge is divided into two parts (blue and red) at a node where the longest side ridge (green line) joins the main one.}
\label{fig4}
\end{figure}

Having decoded the ridge axes in the SRTM grids covering all the mountain ranges of our interest, in the subsequent step we identify the bifurcation points of two or more ridges and associate them with the network nodes of Type 1 (T1). The coordinates of each DEM point being part of a ridge are stored, thus it is straightforward to identify such points: it happens if the same coordinate pair is shared by at least three different segments. We also associate the ridge terminal points with the nodes of Type 2 (T2). Then we connect each node with its nearest neighbours by a binary edge, provided that two nodes are the nearest neighbours if it is possible to move from one node to the other along a ridge axis without passing through any other node. A diagram in Fig.~\ref{fig3}(b) illustrates how a topographic network is constructed from a hypothetical ridge map drawn in Fig.~\ref{fig3}(a). (Note that in the ridge-axis detection procedure we use, the maximum degree possible is $k_{\rm max}^{\rm T}=8$ - see Fig.~\ref{fig1}(b), but this value does not imply that in reality there is no terrain point where more than 6 child ridges detach from their parent ridge). As a result, a tree graph with $N_{\rm T}=N_{\rm T1}+N_{\rm T2}$ nodes and $N_{\rm T}-1$ edges is formed representing the complete ridge structure of a given mountain range. We call this network representation a topographic network since the particular ridges can be easily identified by looking at its topology. The topographic networks for the 9 mountain ranges considered in this work are presented in Fig.~\ref{fig4}.

In order to present the mountains in the ridge network representation, we have to attribute a hierarchy to the ridges. Out of a number of possibilities~\cite{horton1945,strahler1952,hack1957,melton1959,andah1987}, we choose the ridge ordering based on the ridge length. From a network perspective, it is convenient to proxy the metric length of a ridge along its axis with the corresponding network path length. The former is more tedious to calculate since the length of the SRTM grid segments varies from one set of the geographical coordinates to another, while the latter can be calculated straightforward from a network. Such an approach is justified since the side ridges of most ridges are more or less uniformly spaced and the longer the ridge is, the better fulfilled is this correspondence. Therefore, for each mountain range, we calculate the corresponding network diameter $D_{\rm T}$ (i.e., the longest path out of the shortest paths among all node pairs, but for trees this definition is simpler: it is the longest path possible) and identify the main ridge with one of the paths whose length is equal to the diameter. The main ridge has the ridge order $R=1$. Then, for each node of the main ridge, we attribute $R=2$ to the longest network path connecting this node with the most distant $T2$ node in the subnetwork attached to this node. We repeat this procedure with the side ridges of the $R=2$ ridges, attribute them $R=3$, and so on, until all the ridges are ordered. It is easy to see that, for a given mountain range, the highest ridge order possible is $R_{\rm max}=\left \lfloor D_{\rm T}/2 \right \rfloor+1$, where $\lfloor \cdot \rfloor$ denotes floor.

However, it has to be noted that our definition of the main ridge makes it topologically unique with two T2 nodes (as both its ends are unconnected) unlike any other ridge with only a single T2 node. In reality, the main ridge typically coincides largely with the axis of a mountain range that goes roughly parallel to the border of the colliding tectonic plates. Thus, the main ridge length is determined by a particular plate size or a fault length, while the length of its side ridges is typically bounded by a much smaller width of the thrust zone. In many cases this may cause that the main ridge is not only the longest ridge by definition, but also it is incomparably longer than any side ridge (see the main ridges of the Alps and the Himalayas in Fig.~\ref{fig4}, for instance). To make the main ridge resembling topologically other ridges, in the topographic representation we split it into two parts at the node where the longest side ridge is attached. After the splitting we obtain two ridges with a single T2 node each that are always the longest ones in the topographic networks. We let both the parts inherit the $R=1$ order. This division is illustrated in Fig.~\ref{fig4} where the parts of the main ridge are distinguished by different colours (blue and red) together with the longest $R=2$ ridge (green).

\begin{figure}
\begin{center}
\includegraphics[width=0.7\textwidth]{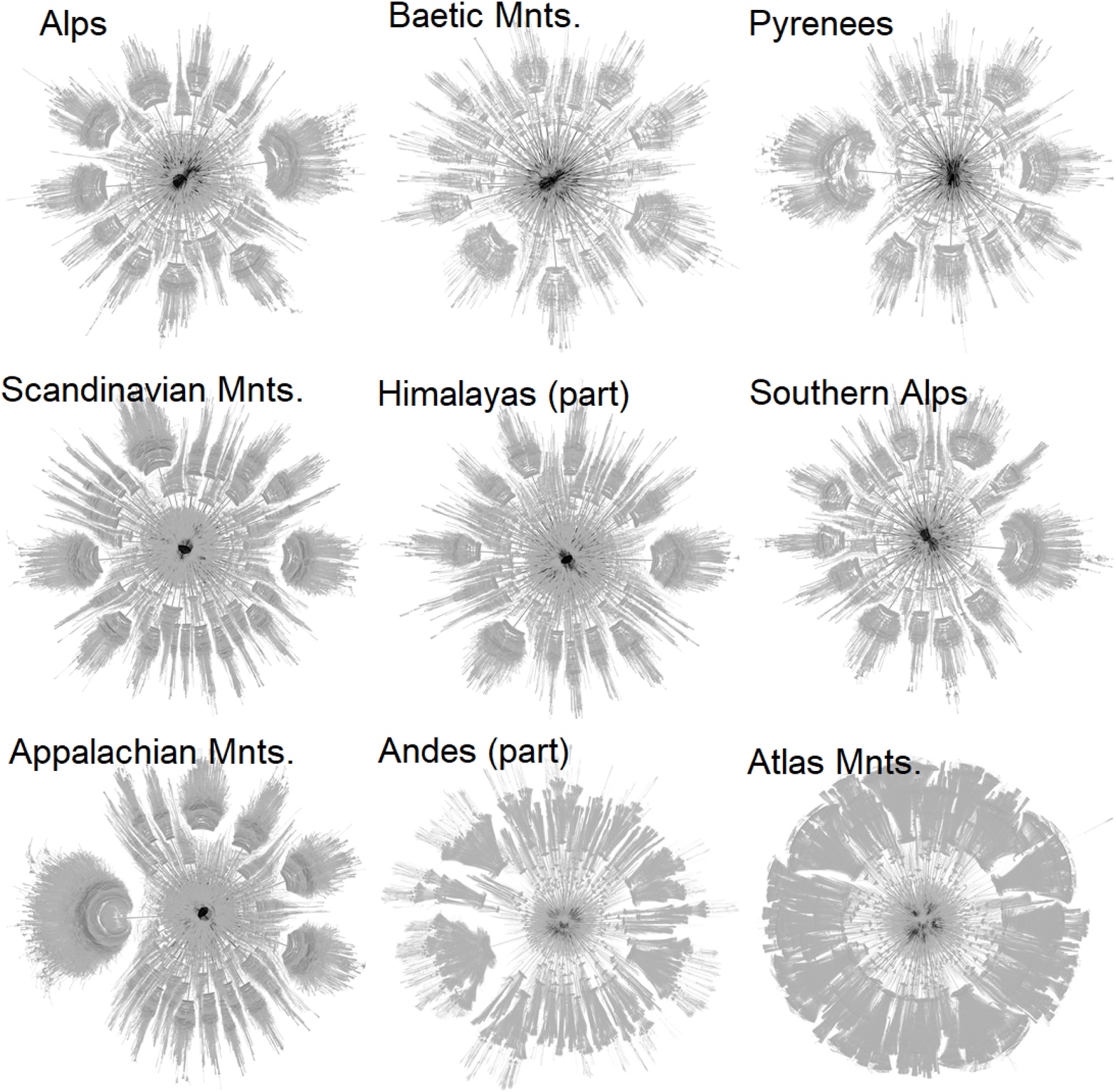}
\end{center}
\caption{Ridge networks for the mountain ranges considered in this work. Each node corresponds to a ridge and each binary edge corresponds to a junction/bifurcation point of the corresponding ridges. Two the most connected nodes for each mountain range represent two parts of a main ridge with $R=1$ (blue and red lines in Fig.~\ref{fig4}).}
\label{fig5}
\end{figure}

In the ridge networks each node represents a ridge and each edge connects two nodes in such a way that the corresponding ridges have a common junction point. As a result, each network of this type has a central hub representing the longer part of the main ridge and a secondary hub representing the shorter part, both linked by an edge. Their remaining nearest neighbours represent the ridges of order $R=2$, their second-nearest neighbours (except for the paths passing through their common edge) represent the ridges of order $R=3$, etc. Fig.~\ref{fig3}(c) shows a ridge network based on a ridge system of Fig.~\ref{fig3}(a). By construction, the number of nodes in a ridge network and the number of the Type-2 nodes in the related topographic network are equal: $N_{\rm R} = N_{\rm T2}$. No similar relation exists that would link $N_{\rm R}$ with $N_{\rm T1}$ except for the inequality: $N_{\rm R} \ge N_{\rm T1}+2$ originating from the fact that two or more side ridges can be connected to their parent ridge at the same point (in fact, about 8\% of the junction nodes have a degree $k^{\rm T} > 3$) and from the fact that in a minimum structure consisting of a main ridge (divided into 2 parts) and a single side ridge there is $N_{\rm T1}=1$ junction node connecting $N_{\rm R}=3$ ridges ($N_{\rm T1}+2$). There is also a condition that relates the maximum ridge order with the diameter of a ridge network: $D_{\rm R} \le 2R_{\rm max}-1$ (the equality holds if a path between two nodes of order $R_{\rm max}$ passes through both $R=1$ nodes). Fig.~\ref{fig5} presents the ridge networks corresponding to the mountain ranges considered here.

\section{Network topology}

\subsection{Node-degree distributions}

By comparing Fig.~\ref{fig5} with Fig.~\ref{fig4}, we see that even though both network representations are acyclic and connected, their topologies are significantly different from each other. The topographic networks are distributed networks with the small maximum node degree $6 \le k^{\rm T}_{\rm max} \le 8 $ and large average shortest path length $461 < L_{\rm T} < 2,435$, while the ridge networks are highly centralized with a clear hierarchy of nodes, the high maximum node degree $852 \le k^{\rm R}_{\rm max} \le 5,330$ and the small-world property: $L_{\rm R} < \ln N_{\rm R}$ (Tab.~2). On the other hand, the different mountain ranges set out in the same representation can show strong similarity irrespective of their type, height, and geographical location. For example, there is a simple relation between the maximum ridge order and the number of nodes in the topographic representation: $R_{\rm max} \approx \ln N_{\rm T}$. It is fulfilled for all the complete mountain ranges analyzed here. The exceptions are the Himalayas and the Andes, only parts of which are considered, and we suppose that their incompleteness might be the cause for the deviations.

\begin{center}
\begin{table}[h!]
\tabcolsep=0.29cm
\begin{tabular}
{|l|l|l|l|l|l|l|l|}
\hline
\multicolumn{1}{|c|}{Range} & $N_{\rm T}$ & $\ln N_{\rm T}$ & $\ln\ln N_{\rm T}$ & $R_{\rm max}$ & $L_{\rm T}$ & $D_{\rm T}$ & $k_{\rm max}^{\rm T}$ \\
\hline\hline
Alps & 510,466 & 13.14 & 2.58 & 13 & 1,252.6 & 3,229 & 7 \\
\hline
Baetic Mts. & 234,117 & 12.36 & 2.51 & 12 & 821.5 & 2,241 & 7 \\
\hline
Pyrenees & 95,356 & 11.47 & 2.44 & 11 & 461.7 & 1,148 & 6 \\
\hline
Scandinavian Mts. & 1,007,154 & 13.80 & 2.62 & 14 & 2,434.4 & 6,574 & 8 \\
\hline
Himalayas (part) & 582,526 & 13.27 & 2.59 & 12 & 1,635.0 & 4,437 & 7 \\
\hline
Southern Alps & 436,955 & 12.99 & 2.56 & 13 & 910.6 & 2,300 & 7 \\
\hline
Appalachian Mts. & 965,668 & 13.78 & 2.62 & 14 & 2,158.1 & 6,232 & 7 \\
\hline
Atlas Mts. & 2,550,922 & 14.75 & 2.69 & 14 & 2,582 & 5,332 & 7 \\
\hline
Andes (part) & 1,008,370 & 13.82 & 2.62 & 12 & 1,857 & 4,382 & 8 \\
\hline
\end{tabular}

\vspace{0.3cm}
\tabcolsep=0.27cm
\begin{tabular}
{|l|l|l|l|l|l|l|l|l|}
\hline
\multicolumn{1}{|c|}{Range} & $N_{\rm R}$ & $\ln N_{\rm R}$ & $\ln\ln N_{\rm R}$ & $D_{\rm R}$ & $L_{\rm R}$ & $\langle m_{\rm R} \rangle$ & $k_{\rm max}^{\rm R}$  & $\beta$ \\
\hline\hline
Alps & 266,540 & 12.49 & 2.53 & 25 & 9.6 & 5.65 & 2,203 & 1.65 \\
\hline
Baetic Mts. & 122,709 & 11.72 & 2.46 & 23 & 8.8 & 5.24 & 1,357 & 1.7 \\
\hline
Pyrenees & 50,027 & 10.82 & 2.38 & 21 & 8.7 & 5.15 & 852 & 1.6 \\
\hline
Scandinavian Mts. & 514,641 & 13.15 & 2.58 & 25 & 9.4 & 5.35 & 4,086 & 1.6 \\
\hline
Himalayas (part) & 304,661 & 12.63 & 2.54 & 23 & 9.1 & 5.50 & 2,286 & 1.7 \\
\hline
Southern Alps & 227,920 & 12.54 & 2.53 & 24 & 9.6 & 5.63 & 1,179 & 1.7 \\
\hline
Appalachian Mts. & 488,822 & 13.06 & 2.57 & 26 & 9.3 & 5.72 & 4,072 & 1.65 \\
\hline
Atlas Mnts. & 1,686,481 & 14.30 & 2.66 & 25 & 8.9 & 5.20 & 5,330 & 1.65 \\
\hline
Andes (part) & 655,892 & 13.39 & 2.59 & 24 & 9.4 & 4.85 & 4,380 & 1.6 \\
\hline
\end{tabular}
\caption{Basic topological characteristics for the topographic (T, top) and the ridge (R, bottom) network representations of all the mountain ranges considered in this work: the number of nodes $N$ (the T1 and T2 nodes are considered together in the T-representation), the maximum ridge order $R_{\rm max}$, the average shortest path length $L$, the network diameter $D$, the mean occupation layer $\langle m \rangle$, the maximum node degree $k_{\rm max}$, and the scaling exponent $\beta$ (the latter was calculated only for the ridge networks). Both the network representations are acyclic and connected.}
\label{si-tab2}
\end{table}
\end{center}

\begin{figure}
\begin{center}
\includegraphics[width=0.5\textwidth]{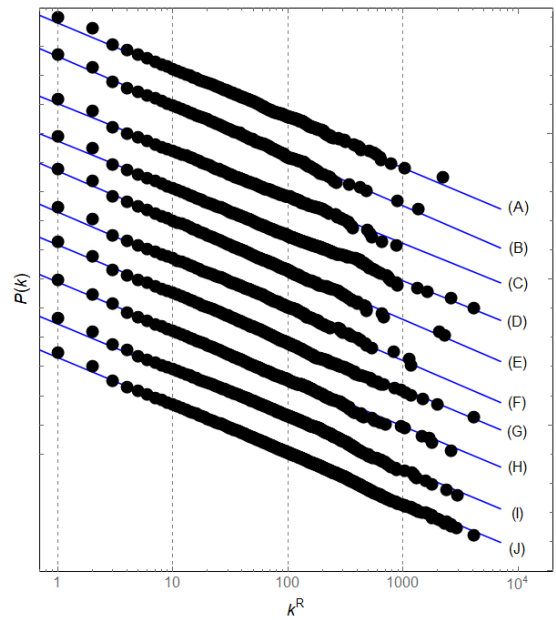}
\end{center}
\caption{The cumulative degree distributions of the ridge networks for different mountain ranges after splitting the main ridge into two parts at a node where the longest side ridge joins the main one (see Fig.~\ref{fig4}) in order to remove its topological distinctiveness caused by the two loose ends. A value of the scaling exponent of the distribution tails is given in each case together with the best-fitted power law function $P(k)\sim ck^{-\beta}$ (straight blue lines, note a log-log scale). The graphs for the individual mountain ranges were shifted vertically in relation to each other for the sake of clarity: (A) Alps, (B) Baetic Mountains, (C) Pyrenees, (D) Scandinavian Mnts., (E) Himalayas, (F) Southern Alps, (G) Appalachian Mnts., (H) the Atlas Mountains, (I) the Andes, (J) all the 9 mountain groups together.}
\label{fig6}
\end{figure}

The most interesting observation regarding the ridge networks comes from Fig.~\ref{fig6} showing the cumulative node degree distributions. It occurs that for each mountain range the best approximation of their shape is a power-law model: $P(X>k)\sim k^{-\beta}$ with the scale-free region spanning 2 or even 3 orders of magnitude $3 \le k^{\rm R} \le k^R_{\rm max}$, depending on the range area and arborescence. (We tested other possible models that resemble partially the scale-free behaviour, like the stretched exponential and the log-normal ones, but the power law model seems to be optimal in this case.) There are some deflections from an ideal power law in some networks for highly connected nodes, but they are rather moderate. For a comparison, we also show the joint cumulative node degree distribution constructed from all 9 mountain groups (Fig.~\ref{fig6}(J)) and observe a similar behaviour to any individual case.

The distribution type observed here is not surprising, however, because the power-law relations are typical for fractal objects the mountain ranges are examples of~\cite{gagnon2006}. What is surprising actually is the values of the scaling exponents $\beta$ that are common across different mountains: $1.6 \le \beta \le 1.7$. As the selected mountain groups differ considerably among themselves in many properties, such similarity might suggest that we observe an effect of the geometric packaging of the ridges within some area, which is governed by the properties of the rock material and the geophysical processes that different mountains are universally subject to, limiting the possible ridge spacing~\cite{pelletier2003,mcguire2013,tucker1998,roering2007}. Thus, we expect that, for different mountain groups that are not included in our study, the node degree distributions for the ridge networks can reveal similar power-law tails.

It is worth mentioning that the pdf scaling exponent $\gamma=\beta+1$ has the values similar ($2.6 \le \gamma \le 2.7$) to many other empirical scale-free networks. For example, there are reports that $2.1 \le \gamma \le 2.7$ for the World-Wide Web (WWW) link networks ($\gamma$ for the outcoming links is larger than for the incoming ones)~\cite{albert1999}, $2.4 \le \gamma \le 2.5$ for the Internet connections at the router level~\cite{faloutsos1999}, $\gamma=2.3$ for the actor co-ocurrence in films~\cite{barabasi1999,amaral2000}, $2.5 \le \gamma \le 3.0$ for the scientific collaboration networks~\cite{newman2001,drozdz2017}, $\gamma=3.0$ for the scientific paper citation networks~\cite{redner1998}, $\gamma=2.8$ for the word-coocurrence networks~\cite{ferrer2001}, $\gamma=2.4$ for the protein interaction networks~\cite{jeong2001}, $2.0 \le \gamma \le 2.4$ for the biochemical cellular pathway networks~\cite{jeong2000}, and $2.4 \le \gamma \le 2.7$ for the currency comovement networks~\cite{kwapien2009,kwapien2012}. It should be kept in mind, however, that such a similarity might be superficial only since the mechanisms leading to the power-laws may be distinct in each case.

\subsection{Fractal analysis of the ridge lines}

\begin{figure}
\begin{center}
\includegraphics[width=0.6\textwidth]{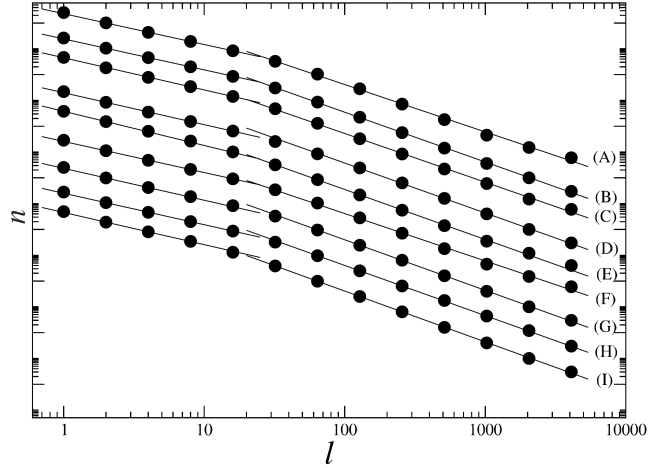}
\end{center}
\caption{The number of occupied boxes $n$ vs. the box size $l$ for the ridge-line maps for different mountain ranges (circles) together with the best-fitted power-law models (straight lines). The plots are shifted vertically for the sake of clarity: (A) Alps, (B) Baetic Mountains, (C) Pyrenees, (D) Scandinavian Mnts., (E) Himalayas, (F) Southern Alps, (G) Appalachian Mnts., (H) the Atlas Mnts., (I) the Andes. In each case, the small-scale power-law regime corresponds to an actual fractal structure, while the large-scale regime indicates a densely populated plane.}
\label{fig7}
\end{figure}

Before we apply fractal analysis to the networks, we consider the ridge-line maps themselves (Fig.~\ref{fig2}(b)) and calculate the box-counting dimension $D$~\cite{mandelbrot1982} for these maps. In order to do this, for each mountain group we divide its area into a number of square boxes of size $l$ and calculate the number of such boxes $n(l)$ that contain a piece of a ridge line. Then, the box-counting dimension $D$ is defined by the following relation:
\begin{equation}
n(l) \sim l^{-D}
\end{equation}
if such a relation holds for a significant range of $l$. Typically, if $D$ is a non-integer number, a box-covered object is fractal. Fig.~\ref{fig7} presents plots of $n(l)$ for the boxes of length $1 \le l \le 4096$ data points (equivalent to $\sim 90$ m $\div \sim 368$ km along a meridian). There are 2 scale ranges where the dependence is approximately linear (i.e., power-law actually): the small-scale one ($1 \le l \le 16$) with $1.2 \le D \le 1.3$ and the large-scale one ($32 \le l \le 4096$) with $1.8 \le D \le 2.0$ (Tab.~3). By looking at Fig.~\ref{fig2}(b), we see that the appearance of these two regimes is related to the different small- and large-scale properties of the ridge-line maps. For the scales less than 2 km, which correspond to typical inter-ridge distances, the box-counting algorithm detects the actual fractality related to the dendritic structures. However, since the ridges are close to each other, above the scale of 2 km almost any box contains some ridge-line part; the topology ceases to be fractal and becomes just a plane. This is why $D$ is so close to 2 in this case. We conclude that the ridge-line structure of any mountain range is fractal below 1-2 km and space-filling above it. Nevertheless, this does not exclude the obvious visual fractality of the mountain ridges on any scale because such a large-scale self-similarity is encoded in elevations rather than in the mere ridge axes (compare both parts of Fig.~\ref{fig2}). Please note that the river networks also reveal two types of geometrical behaviour with the fractal dimension $1.1 \le D \le 1.2$ below a cetrain scale and $1.8 \le D \le 2.0$ above it~\cite{grey1961,mandelbrot1982,hjelmfelt1988,tarboton1988,labarbera1989}.

\begin{center}
\begin{table}[h!]
\begin{footnotesize}
\begin{tabular}
{|l|c|c|c|}
\hline
\multicolumn{1}{|c|}{Range} & $D_{\rm small}$ & $D_{\rm large}$ & $d_f$ \\
\hline\hline
Alps & 1.20 & 1.84 & 1.65 $\pm$ 0.11 \\
\hline
Baetic Mountains & 1.20 & 1.93 & 1.70 $\pm$ 0.13 \\
\hline
Pyrenees & 1.25 & 1.89 & 1.61 $\pm$ 0.14 \\
\hline
Scandinavian Mountains & 1.24 & 1.94 & 1.62 $\pm$ 0.12 \\
\hline
Himalayas & 1.32 & 1.94 & 1.68 $\pm$ 0.12 \\
\hline
Southern Alps & 1.21 & 1.84 & 1.68 $\pm$ 0.11 \\
\hline
Appalachian Mountains & 1.24 & 1.96 & 1.65 $\pm$ 0.10 \\
\hline
Atlas Mountains & 1.25 & 1.97 & 1.65 $\pm$ 0.10 \\
\hline
Andes & 1.22 & 1.95 & 1.66 $\pm$ 0.11 \\
\hline
\end{tabular}
\caption{The box-counting dimension calculated for the ridge-line maps (the small-scale $D_{\rm small}$ and the large-scale $D_{\rm large}$ values) and the network box-covering dimension $d_f$ calculated for the topographical networks.}
\end{footnotesize}
\label{tab3}
\end{table}
\end{center}

\subsection{Fractal analysis of the networks}

As the mountain ranges have fractal structure, we are curious whether this property can also be observed in their network representations. These representations are defined in an abstract space, in which the position of a given node is determined by its relation to other nodes in the ridge configuration rather than by any system of spatial coordinates. Therefore, the most adequate method of quantifying network fractality in this case is the renormalization-based approach called the box-covering~\cite{song2005}. 

In the box-covering method a network is divided into node clusters (``boxes'') and subsequently coarse-grained on various ``length'' scales. One first defines the distance parameter $l$ that bounds the path length between the nodes belonging to the same cluster. Next, a seed node that is considered as a center of the first cluster is randomly chosen. Then, the number $N_c$ of the clusters is calculated after partitioning the network into the node clusters in such a way that the minimum path length between the nodes belonging to the same cluster is not longer than $l-1$. The same is done independently for different draws of the seed node and the average number of clusters $\langle N_c \rangle$ is determined. In a subsequent step, the renormalization, the clusters are replaced by single nodes (linked if there existed at least one connection between the nodes belonging to the corresponding clusters) and a new network consisting of such nodes is formed and subject to the analogous partitioning and replacement procedure. These steps can be repeated until the network is reduced to a single cluster. However, since the consecutive renormalized networks have similar topological properties to the original network (for example, the node degree distributions show the same scale-free behaviour), it is sufficient to calculate $\langle N_c(l) \rangle$ for different values of the parameter $l$. The network is fractal if the following power-law relation holds:
\begin{equation}
\langle N_c(l) \rangle / N \sim l^{-d_f},
\label{eq::power-law}
\end{equation}
where $\langle N_c \rangle$ is the average number of the network clusters, $N$ is the number of the network nodes, and $l$ is the distance parameter. In this case the parameter $d_f$ is considered the fractal dimension of the network. It was documented in literature that the empirical networks can show either the fractal scaling~(\ref{eq::power-law}) or a non-fractal behaviour of $\langle N_c(l) \rangle/N$, e.g., its exponential decay~\cite{song2005}.

\begin{figure}
\includegraphics[width=0.5\textwidth]{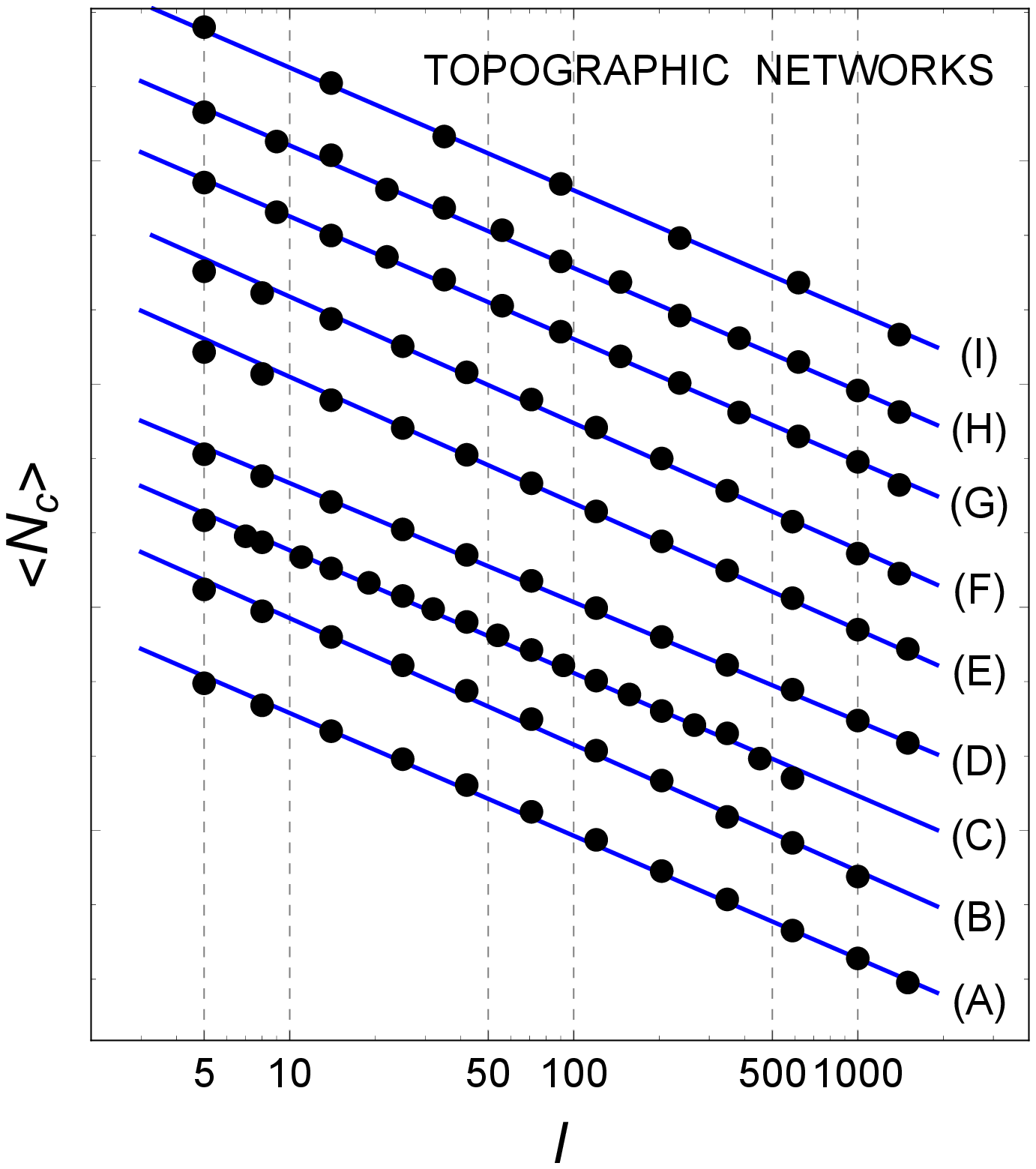}
\includegraphics[width=0.5\textwidth]{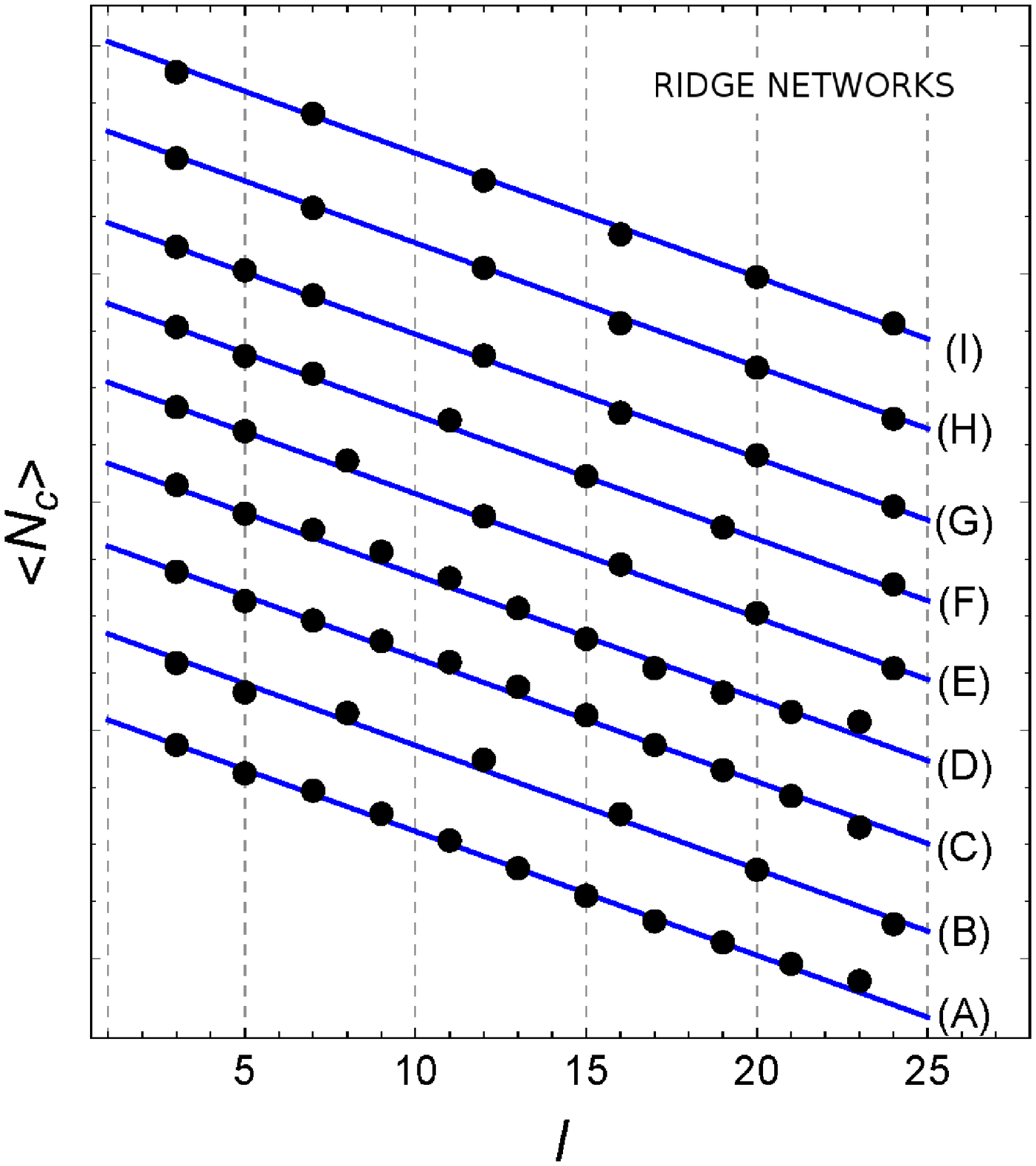}
\caption{The average number of clusters $\langle N_c(l) \rangle$ after partitioning the topographic networks (a) and the ridge networks (b) for different mountain ranges in such a way that the minimum path length between any two nodes belonging to the same cluster is less than $l-1$ (black circles). The least-square fits of the power-law function $f(l) = c l^{-d_f}$ (a) and the exponential function $g(l) = c \exp(-l/2)$ (b) are shown by the straight blue lines (note a log-log scale in (a) and a semi-log scale in (b)). The graphs for the individual mountain ranges were shifted vertically in relation to each other for the sake of clarity: (A) Alps, (B) Baetic Mountains, (C) Pyrenees, (D) Scandinavian Mnts., (E) Himalayas, (F) Southern Alps, (G) Appalachian Mnts., (H) the Atlas Mnts., (I) the Andes.}
\label{fig8}
\end{figure}

Here we calculate the fractal dimension $d_f$ of the network for both the topographic and the ridge network representations. The results for the topographic networks showing the scale-free dependence with the scaling exponents $1.6 \le d_f \le 1.7$ for $l>50$ being similar for each mountain group are displayed in Fig.~\ref{fig8}(a). In contrast, the ridge networks do not present the power-law dependence~(\ref{eq::power-law}). Instead, the exponential relation $\langle N_c(l) \rangle \sim e^{-l/2}$, where $\langle N_c \rangle$ is the average number of the network clusters, can approximate the empirical data for all the values of $l$ in this case (Fig.~\ref{fig8}(b)). It is interesting that although each representation has different properties: fractal in the case of the topographic networks and non-fractal in the case of the ridge networks, different mountain ranges show similar behaviour if the same representation is considered. These results are not surprising, though. On the one hand, as the topographic networks inherit basic features of the ridge structure of the real mountains, the network fractality comes as its consequence. On the other hand, the scale-free character of the ridge networks implies the existence of massive hubs, and this, by forming highly populated node clusters, limits the total number of clusters in the network even for the moderate values of $l$ and produces the exponential decay of $\langle N_c(l) \rangle$~\cite{song2005,song2006}.

The box-counting dimension $1.2 \le D \le 1.3$ of the ridge maps is significantly lower than the network box-covering dimension $1.6 \le d_f \le 1.7$ for the same mountain groups. To understand this relation one has to realize that on the short scales corresponding to the fractal ridge-line structures, the box-counting algorithm sees predominantly the 1-D line segments, while the box-covering algorithm sees rugged topographic networks, where any node is a branching point by definition. Therefore the network dimension $d_f$ must be substantially higher than the map dimension $D$, although both are topologically restricted: $d_f,D \in [1,2]$. It should be noted that $D$ and $d_f$, while both describe the fractal structure of the mountain ridges, are rather unrelated with each other. This is because $D$ measures how the ridge lines fill the available space with relatively little attention to ridge branching, while $d_f$ measures how strongly the ridges are branched and how many side ridges leave the same branching point with no attention to the ridge line shape and space filling.

\subsection{Multifractal network analysis}

As the rugged terrain surface shows signatures of multifractality~\cite{gagnon2006,lovejoy2007,schertzer2011}, we then addressed a question whether the morphological differences between mountain ranges can manifest themselves in the topographic networks on the multifractal level. A multifractal can be viewed as an object composed of a number of convoluted fractals that have different fractal dimensions (monofractals), and as a consequence it is much complex than its constituents. In our present case, a multifractal network may have different fractal topology in its different parts. We perform an analysis that is able to reveal such a structure, if present, in the topographical networks for some mountain groups.

An algorithm that is commonly used in multifractal network analysis is the sandbox algorithm~\cite{tel1989,liu2015,rendon2017,mali2018}. First, one has to create a distance matrix for a network by calculating the shortest path (for a tree the only path) lengths between all the nodes. The algorithm starts by chosing randomly a seed node that serves as a sandbox centre. A sandbox has a variable radius $r$ and one counts the number of nodes $n$ that fall in the sandbox for a particular value of $r$ in order to work out a functional dependence $n(r)$. This is repeated for a sufficient number of different seed nodes.

The generalized fractal dimension $D_q$ with the R\'enyi parameter $q \in \mathcal{R}$ can empirically be derived by applying the following formula:
\begin{equation}
D_q = \lim_{r \to 0} { \ln \langle [n(r) / N]^{q-1} \rangle \over (q-1) \ln (r/d) },
\end{equation}
where $N$ is the total number of nodes in the network and $d$ is the network's diameter, which plays a role of a normalization constant. The brackets $\langle \cdot \rangle$ donote averaging over the seed nodes. The dimensions $D_q$ can be estimated by a linear regression of $\ln \langle [n(r)]^{q-1} \rangle$ against $(q-1) \ln (r/d)$ if the data allows for such a regression to be applied. If the regression-line slopes decrease with increasing $q$, one may conclude that the network under study is multifractal, while a lack of such variability suggests a monofractal structure.

\begin{figure}
\includegraphics[width=1.0\textwidth]{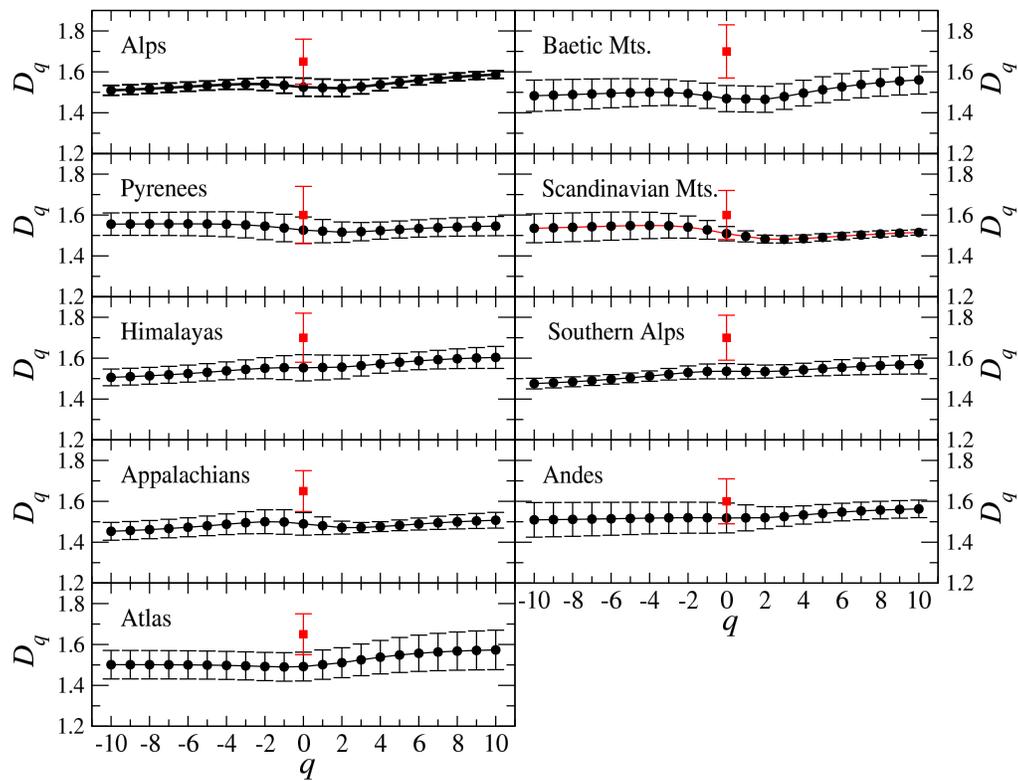}
\caption{Generalized fractal dimensions $D_q$ calculated for the topographic networks with a multifractal sandbox method (black circles) and the fractal dimension $d_f$ calculated for the same networks with the box-covering method (grey squares). Each panel presents results for the network representing a particular mountain range. The error bars denote the standard deviation derived for 10 independent choices of a seed-node set (each one consisting of 50 nodes) in the sandbox method or for different regression-line fits in the box-covering method. The increasing parts of the $D_q$ plots are a computational artifact related to finite-size effects.}
\label{fig9}
\end{figure}

We applied the above algorithm to the topographic networks and obtained the results collected in Fig.~\ref{fig9}. A range of the $D_q$ variability for $-10 \le q \le 10$ is $|\Delta D_q| \le 0.1$, which in principle might indicate weak multifractality. However, our experience with multifractal analysis (see e.g.~\cite{drozdz2009}) suggests that such a small difference can entirely be attributed to finite-size effects. Then, accordingly, the topographic networks may be considered monofractal. This conclusion is supported by an observation that the $D_q$ dependence is not monotonous and in each case there is a range of $q$ for which it is an increasing function. As this behaviour is not possible in theory, its existence comes solely from the statistical uncertainty of the results. Thus, since the finite-size effects' influence is of the same size as $\Delta D_q$, the monofractality hypothesis better complies with the results than the weak-multifractality one.

The fractal dimension of the networks obtained with the sandbox approach is $1.45 \le D_q \le 1.60$ with $1.5 \le D_0 \le 1.55$ (see the circles in Fig.~\ref{fig9}), which is slightly less than the dimension $d_f$ obtained with the box-covering method (the squares in Fig.~\ref{fig9}), which is a typical relation for both methods, but if one takes the statistical uncertainty into account and adds the error bars, it becomes evident that in fact both methods offer largely overlapping outcomes.

Our results show that different mountain ranges can be distinguished based on neither the fractal nor the multifractal network approach. This implies that the ridge-line structure for any mountain range is universal and reflects more the drainage structure, fluvial and gravitational erosion, and some related physical optimization processes, which are similar in different parts of the world, than the tectonic and morphological properties of that range. The latter are encoded in the elevation distribution, which we do not investigate here. We also do not expect that, for instance, applying a different ridge ordering scheme can cause any distortion of the observed universality as the same ordering schemes must be used systematically across the data. However, some small differences in the numbers, like the power-law exponents, may occur, indeed.

\section{Summary}

In this work, we considered the ridge structure of several mountain ranges located on different continents in two distinct network representations: the topographic representation (in which the ridge junctions and the ridge ends were depicted by nodes and the edges linked those nodes whose corresponding terrain points were directly connected with a ridge) and the ridge representation (in which the ridges were depicted by nodes and the ridge junctions were depicted by edges). We found that the topographic networks are distributed networks with the long average shortest paths and a fractal structure with the fractal dimension $1.6 \le d_f \le 1.7$. This can be compared with the box-counting dimension of the ridge-line structure: $1.2 \le D \le 1.3$. Despite these numerical differences, the fractal network structure is likely inherited directly from the fractality of the ridge lines. In contrast, the ridge networks do not possess any fractal structure owing to their assortativity: the hubs are connected to other hubs directly without any relaying nodes in between. However, they are scale-free with the power-law exponent $1.6 \le \beta \le 1.7$.

We expect that although we chose here a specific network representations, which are distinct from the standard terrain-to-graph mappings applied in geoscience, the main outcome of this work (the geographical universality) would not have altered if we had based our study on, e.g., the surface networks. If we abandon the node spatial coordinates of the surface networks and consider the internode connections only, the topographic networks may be viewed as subgraphs of the surface networks. The peaks, which belong to a set of the surface network nodes, are typically the points where side ridges start their course from, so the peaks are also the topographic network nodes. Even though the surface networks comprise not only the ridges, but also the channels, we do not expect that this can destroy the universality of outcomes as, to a large extent, the channel systems mirror topologically the ridge systems~\cite{werner1988}.

It is interesting that all the above results are similar for all the considered mountain ranges irrespective of their location, height, area, age, and morphology, which even carries a signature of universality. This observation is especially counter-intuitive if one takes a look at the topography of such ranges as the Alps and the Appalachian Mountains. The former consist mainly of the dendritic systems of ridges and valleys, while the latter in a large part have a form of the long, alternating ridges and valleys with a trellis-like drainage structure. Despite such a structural variability, the topographic networks for different ranges cannot be distinguished even by a multifractal approach. We conclude that the ridge lines are distributed roughly homogeneously across the mountaineous terrain.

We believe that a source of this structural universality can be sought in the universality of the mountain-shaping processes, like the fluvial erosion, and from the universal properties of the rocks that the mountains are built of, which set limits on the possible hillslopes, the distances between the neighbouring ridges, and the ridge lengths. As earlier studies showed, these factors combined together may lead to characteristic optimization of any ridge structure, e.g., in order to minimize a surface-to-volume ratio~\cite{mark1981}. The observed universality may be a consequence of such optimization.

In this context, it has to be stressed that the real-world ridge formation processes are far more complicated than the model mechanisms of the fractal network growth~\cite{song2006}, thus one cannot infer much information on such processes by studying the related available models. Our results open space and indicate a need for introducing a new class of growing network models that are able to mimic the ridge and valley formation and channel rewiring since the very beginning of their existence on some territory to a mature dendritic form. Such models may shed new light on the mountain structure formation and evolution; they might also be interesting from a graph-theory point of view.

We suggest also a possibility to use the ridge network representation as an aid in cartographic generalization since such a representation is conveniently compact and allows one to identify easily both the nodes/ridges of significant centrality (to be preserved) and the peripheral ones of little topological significance (to be neglected). Further, if applied to channels instead of ridges, the channel representation (i.e., an analogue of the ridge representation) can be convenient for a fast assessment of a drainage basin area and flow volume of a particular channel at its mouth in relation to other neighbouring channels as these quantities are closely related to the corresponding node's centrality (roughly, the higher it is, the larger the basin area and the flow volume is). The same refers to a watershed's significance: the higher a node's centrality, the more important a watershed defined by the corresponding ridge is. Such assessments are less straightforward if one considers the topographical and surface networks, for instance.

We expect that much useful information can additionally be inferred from a related analysis based on weighted network representations. Specifically, by including the edge weights, one can observe a richer multifractal structure as multifractality might be contained in the ridge heights (and the heights can be used for constructing weighted networks). For example, it is interesting to look for possible correlations between multifractality and mountain relief.

\section*{Acknowledgements}

Research supported by the National Science Centre (Poland), grant no. 2017/01/X/ST2/00020.

\end{document}